\documentstyle[prl,aps,epsf]{revtex}

\newcommand{\eq}[1]{\begin{equation}\label{#1}}
\newcommand{\eqs}{\begin{equation}}
\newcommand{\en}{\end{equation}}
\begin{document}

\twocolumn[{

\title{Invading interfaces and blocking surfaces in high
dimensional disordered systems}
\author{Omri Gat and Zeev Olami}
\address{Chemical Physics Department, Weizmann Institute of Science \\
Rehovot, Israel 76100\\\today}
\maketitle
\widetext
 \leftskip 54.8pt
 \rightskip 54.8pt
\begin{abstract}
We study the high-dimensional properties of an invading front in a
disordered medium with random pinning forces. We concentrate on
interfaces described by bounded slope models belonging to the quenched
KPZ universality class. We find a number of qualitative transitions in
the behavior of the invasion process as dimensionality increases. In
low dimensions $d<6$ the system is characterized by two different
roughness exponents, the roughness of individual avalanches and the
overall interface roughness. We use the similarity of the dynamics
of an avalanche with the dynamics of invasion percolation to show
that above $d=6$ avalanches become flat and
the invasion is well described as an annealed process with correlated
noise. In fact, for $d\geq5$ the overall roughness is the same as the
annealed roughness. In very large dimensions, strong fluctuations begin
to dominate the size distribution of avalanches, and this phenomenon
is studied on the Cayley tree, which serves as an infinite dimensional
limit. We present numerical simulations in which we measured the
values of the critical exponents of the depinning transition, both in
finite dimensional lattices with $d\leq6$ and on the Cayley tree,
which support our qualitative predictions. We
find that the critical exponents in $d=6$ are very close to their
values on the Cayley tree, and we conjecture on this basis the
existence of a further dimension, where mean field behavior is obtained.
\end{abstract}
\vskip 5mm
}]

\narrowtext
\section{introduction}
The problem of interface growth and wrinkling in disordered and noisy
systems has been the subject of much interest~\cite{gen}.
A special attention was given lately to interface growth in disordered
systems. Examples of such systems are two fluid displacement flow in
porous media\cite{exp1}, invasion of water into paper\cite{dis1} and
magnetic domain movement in spin systems with quenched disorder.  The
distinction between such disordered systems, where the noise is
quenched, and
cases were the motion is induced by random annealed noise has been an
important part of this study\cite{gen,gen2}. It is known for some time
that the critical properties of interfaces in the two kinds of systems
are different ~\cite{gen,dis1,gen2,dis}.  As a consequence of the time
independence of
disorder in a quenched system an interface invading a
quenched medium may be pinned, if the driving force is smaller than
some threshold value, while in the annealed version it will always
grow. The depinning transition is a critical phenomenon, in which the
correlation length diverges as one approaches the threshold driving
force value.  Thus, there is usually a marked distinction between the
behavior of the quenched noise problem and the annealed versions, and
they define different universality classes.

Advancing interfaces are usually characterized by a roughness exponent
$\chi$ relating the average width of the interface $W$ to the system 
size $L$,

\eq{chi}W=L^\chi.\en

In most cases, the interface has also a self-affine structure characterized by
the same roughness exponents.
A particular difference between annealed and quenched growth is that the
roughness exponent $\chi_q$ of an interface growing in quenched
disorder is generally different from
$\chi_a$, the roughness of the annealed counterpart.    
An important example is bounded slope growth, which may be
described by the KPZ equation \cite{kpz} for the annealed case~\cite
{gen}. The quenched model \cite{gen,dis,snne,opz2} has a
different roughness.

An interesting aspect of quenched growth processes is the mapping
which exists between such problems and the percolation of random
surfaces. In fact, a pinned interface describes a blocking surface which
traverses from one side of the system to the other. If the interface
is depinned, on the other hand, no such blocking surface
exists. In the simplest example of 1+1 dimensions, the blocking
surface describes a line in a directed percolation
cluster\cite{dis1,dis}. In other
dimensions such generalized percolation processes are less well
understood. Thus, we can identify the directed percolation transition
with the depinning transition in 2 dimensions.

We want to emphasize that the analysis presented in this paper relates
to a universality class which is realized in many different
contexts. It was found in studying other systems, that much useful
information and insight can  be gained by studying the behavior in
spaces of varying dimensionality. This is the case also for the
interface growth model studied here. Since the model is rather
involved we discover a number of qualitative transitions as the
dimensionality increases.

Turning back to the dynamics of interfaces we note that critical
behavior in such interfaces can be observed using two different
driving mechanisms. First, one observes critical properties near the
threshold force level  $p_c$, required for the interface depinning.
Alternatively, one may drive the system at the point of weakest resistance
using a constant infinitesimal current. This driving causes the system to
self-organize into a critical state. The critical interface in both cases
is that of a self-affine interface with infinite range correlations.
Interfaces produced  by the two methods are intimately related.  We
will use both methods in our discussions.

Analyzing interface growth driven by infinitesimal current, we can
define a key concept for the understanding of such processes, namely
the associated process (APs) \cite{opz2,opz1}. Associated processes are
defined as the domains covered by the growth at periods during which the
resistance of the medium is below a certain threshold value. 
In annealed disorder one cannot define such processes and this is the
main source for the distinction between annealed and quenched growth models. 
We analyze the statistical properties of the APs, and in particular
their accompanying roughness exponent $\chi_c$
which is defined by the scaling of the height of the APs with respect to their
lateral dimension. We stress that $\chi_c$ is an independent exponent
and may be different from $\chi_q$, since the latter is created by a
process of stacking of independent critical APs.
In fact, $\chi_c$ is only 
a lower bound for the overall roughness $\chi_q$
\cite{opz2}. Thus in general, blocking surfaces have two
independent roughness exponents. The first is the global one, and the
second is the scaling  exponents of the voids between the surfaces.

In a previous letter \cite{letter} 
we extended the discussion and
analysis of interface growth as a series of APs to an arbitrary number
of space dimensions, and used the AP concept to derive a scaling theory
of interface growth in all dimensions. 
We argued from general principles
that the APs become less rough as the dimensionality increases and
eventually become flat objects at a finite dimension $d_c$.  For space
dimensions $d>d_c$ the APs are fractal. Note that this dimension is
critical in the sense that the scaling relation change qualitatively
reflecting a fundamental change in the dynamical process. However a
second critical dimension can exist where the critical behavior
will be the same as in the infinite dimensional limit. 

In this paper we augment the
discussion in~\cite{letter} in several ways. 
We consider the dynamical process that occurs during the evolution of
a critical
AP. The set of sites which are active at a certain instant is
sparse. Therefore we  conjecture that it is a  percolative dynamical
process. Using this claim we can put limits on the maximal possible AP
roughness $\chi_c$ as a function of $d$ and we are able to show that
$d_c=6$ for the quenched KPZ universality class. This is related to
the observation in \cite{per} that the dynamical exponent $z$ equals 2
for $d\geq6$.

We next present simulations of the model on lattices of up to six
dimensions, verifying that
 the $d_c$ is indeed equal to 6.  We analyze the the overall
interface growth and show that the interface  roughness
exponent, $\chi_q$ are the same as
the exponents $\chi_a$ of the annealed KPZ equation above $d=4$. 

The infinite dimensional limit of this problem can be analyzed using
a realization of the model on a Cayley tree
 with an additional height coordinate. Section three is devoted to
this analysis. We performed simulations on this model and found two
main outstanding features. The first one is the existence of strong
anomalous behavior, resulting  in power law distribution functions
with changing exponents in the subcritical range. Secondly,
near the depinning transition a critical behavior emerges  with
a strong anomalous background. The values of the scaling
exponents seem to be the proper infinite dimensional limit of the
model. In fact the scaling exponents in six dimension, which is the
highest dimensional system we analyzed numerically, are quite close to
the results on the Cayley tree. This is an indication that the
critical dimension of the problem, where all exponents reach their
infinite dimensional limit is close to~6.
     
In section 3 we discuss two
variants of the model on the Cayley tree. The first is a
straightforward generalization of the bounded slope model
we just described, in which the bounded slope condition permits
avalanches propagating in all directions,
allowing for self-interaction by a re-activation of a
site by a daughter site. This self-interaction prevents us
from obtaining
a closed form solution. Hence we define a simplified
version in which the bounded slope condition is enforced only in
a preferred direction. This directed version is
simplified enough that analytic calculation can be done, using a
mapping to an infinite state branching process. However, the directed
version is not a satisfying high-dimensional limit, since the
anomalous fluctuations mentioned above swamp out the critical behavior
of the transition, and do not allow for an explicit calculation of
scaling exponents.

\section{The invasion process and interface structure in high dimensions}
\subsection{Basic definitions and scaling relations}
The discrete Buldyrev-Sneppen model \cite{dis1,dis,snne}, which
belongs to the
quenched KPZ universality class, was originally defined on a 1+1
dimensional lattice. At each site there is a quenched random pinning force 
$f(s)$ distributed uniformly between 0 and~1. There are two related
ways to define the invasion process. In the constant current version
one has to locate the site of minimal $f$ value on the interface. This
site is activated by increasing the interface height at that
point. Next one checks whether the activation of this minimal site
creates a local slope which is larger than a threshold value. If such
an event occurs, the neighboring site is also activated to reinstate a
sub-threshold local slope. These two operations are carried out
repeatedly creating an advancing interface. Another algorithm for
advancing the interface is the activation of all sites on the
interface whose value of pinning force $f$ is smaller than a fixed
driving force value $f_0$, followed by avalanches to enforce the
bounded slope condition, and repeating these operation. If $f_0$ is
chosen such that it is smaller
than a critical value $p_c$, the interface will be pinned eventually,
and if $f_0>p_c$, the interface will continue to propagate indefinitely.

A very fruitful way of describing the invasion process by constant
current is to introduce the concept of associated
processes \cite{opz2,opz1}. They are defined by examining the value
of $f$ for each
activated site. An $f_0$ AP is a set of activations with $f$ values
below the threshold value $f_0$.  An activation above $f_0$
concludes  an $f_0$ AP and starts a new one. The
APs of $f_2$ may comprise several $f_1$ APs if $f_1<f_2$. The
largest APs are the $p_c$ APs. In the constant current algorithm, no
site with $f>p_c$ can be activated. An alternative way to construct an
AP, is to switch to the constant force $f_0$ algorithm immediately
after updating a site with $f>f_0$. The process will be blocked, and the
cluster which is created is precisely an AP, and this fact relates the
two driving mechanisms.

The definitions of the model and the two driving methods can be
carried through unchanged to lattices of higher dimensionality. One
defines an interface height coordinate in each site, and performs the
invasion process according to the rules just describes. The lattice
does not have to be an Euclidean lattice, and we will consider
interfaces growing on a Cayley tree in section 3.

For $f_0$ close to $p_c$ the APs display critical behavior, where 
$\Delta f\equiv p_c-f_0$ measures the closeness to criticality of the
AP. We are interested in the distribution of AP sizes (or masses) $s$
defined as the number of activations in the AP, and in the RMS
lateral dimension $r_\parallel$. 
The following scaling forms are known to be valid for $d$-dimensional
APs near $f_c$:
\eq{sc-def} p(s)=s^{-\tau}g(s/\Delta f^{-\nu}),\quad
p(r_\parallel)=r_\parallel^{-\tau_\parallel}g_\parallel(r_\parallel/\Delta
f^{-\nu_\parallel}),\en
and the fractal dimension $D$ of the APs is defined by
\eq{sc-frc}s\sim r_\parallel^D. \en
The scaling relations
\eq{sc-par} \nu_\parallel=\nu/D\qquad\tau_\parallel-1=D(\tau-1)\en
follow immediately from the definitions (\ref{sc-def}) and (\ref{sc-frc}).
If $D>d$, the APs are rough objects, that is, they have a width
$r_\bot$ related to $r_\parallel$ by
\eqs r_\bot\sim r_\parallel^{\chi_c}, \en
where $\chi_c=D-d$.

An exponent relation 
proved in \cite{opz1} using the properties of the model is
\eq{sc-opz} 1=\nu_\parallel(d+1-\tau_\parallel). \en
This relation reduces the number of independent static exponents to~2
and remains valid as long as the AP roughness
exponent $\chi_c$ is positive.
A  relation equivalent to $(\ref{sc-opz})$ is
\eq{gamma} \gamma=1+\nu\chi_c, \en
where $\gamma= \nu (2-\tau)$, is the scaling exponent of the average
cluster size with respect to $\Delta f$.

We will show below that there exists a finite dimension $d_c$ such that
 $\chi_c=0$ when $d>d_c$.  
Above $d_c$ the scaling relation (\ref{gamma}) becomes simply
$\gamma=1$. However, the other exponents and
the fractal dimension continue to depend on the dimension, and
therefore $d_c$ does not mark the transition to mean field behavior.

The geometric structure of an AP is qualitatively different in $d>d_c$.
Studying the rules of the model indicates that whenever $\chi_c$ is
positive, the APs cannot be fractal, since rough
clusters have a typical width $r_\bot$, which through
avalanches prevents the formation of holes with diameter smaller than
$r_\bot$. For $d\leq d_{c}$ the APs may be multiply connected, but
not fractal. Above $d_{c}$, since $D\leq d$ the AP become objects with
order 1 thickness, which may  fractalize  by having  holes of any size. 

\subsection {Three types of roughness exponents}
In the previous subsection we defined the roughness exponent $\chi_c$
which characterizes the statistics of the APs. It is important to
realize that $\chi_c$ is not necessarily equal to the exponent $\chi_q$
defined in eq. (\ref{chi}), which describes the interface as a whole.
However, a simple consideration shows that $\chi_q$
cannot be smaller than the value of $\chi_c$. In fact, from time
to time an AP is generated which encompasses the whole system, and
then the entire interface identifies with a critical AP (see also \cite
{opz2}). 
The interface growth is also related to the annealed process, since
the stacking of APs is similar in spirit to the process described by
annealed equations. This issue is dealt with in more detail below, but
here we would like to note that it follows from our discussion that
the roughness of the annealed dynamics $\chi_a$ is also a lower bound
for $\chi_q$, so we may write
\eq{bounds}
\chi_q \geq \max(\chi_{a},\chi_c).
\en
Since the quenched growth includes a non-trivial interaction between
the processes defining
$\chi_{a}$ and $\chi_c$ its roughness can
actually be larger than both $\chi$'s. 

As an example one can consider the $1+1$ 
dimensional case. Since the APs are related to
directed percolation, their roughness is $\chi_c =0.63$\cite{dis1,dis}. The
roughness for  the annealed dynamics is $\chi_a=0.5$, so that $\chi_c$
dominates $\chi_a$. In fact, it has been found (see \cite{opz2}) that in
this case the
system roughness $\chi_q$ is slightly larger than $\chi_c$. In higher
dimensional systems $\chi_c$
decreases, and the difference between $\chi_q$ and $\chi_c$ becomes more
pronounced.

As an application of this distinction, we note that when constant
force driving is used, starting from a {\it flat} interface as 
initial conditions, the created clusters will have roughness $\chi_q$
rather than $\chi_c$. In fact the clusters will consist of many APs
since the initial surface cuts through them. This difference will be
very manifest in high dimensions, where the APs are flat and fractal
whereas the constant force clusters are rough \cite{shechter}.

\subsection{Dependence of $\chi_c$ on dimensionality}
It is apparent that growth in APs can happen in two ways; either
upward by activations, or sideways via avalanches.
The sideways action becomes more
important as the dimensionality increases, and $\chi_c$ therefore
decreases as a function of $d$. A naive expectation based on this 
scenario is that in sufficiently
high dimension the sideways growth becomes non self-interacting. If
this argument were true, the APs should belong to the percolation
universality class in large enough dimensions. This simple behavior
does not occur, since the bounded slope condition induces
self-interaction which does not decrease when $d$ increases, as we
explain in detail in the next section where we analyze the model on
the Cayley tree.

There is however a subtle connection between the dynamics of a growing AP and
the dynamics of a percolation cluster. In the following subsection we
exploit this analogy to argue that in some finite dimension $d_c$ the
APs become flat objects with $\chi_c=0$. As explained above, $d_c$
does not mark the transition to mean field behavior. Some exponents,
such as the AP fractal dimension $D$,
change as a function of $d$ above $d_c$. Hence, for $d>d_c$ the APs
are fractal objects with $D<d$ (see subsection~2.1). We expect that
$D$ saturates at some finite value $D_\infty$, as indicated by the 
Cayley tree analysis presented in the next section.

\subsection{Dynamical scaling and critical dimension}
The discussion presented so far has concentrated on the AP static
exponents such as $\tau$ and $D$. It is interesting that
considerations involving dynamic exponents provide further information
on static exponents as well. 
Although the arguments given here are not rigorous, we believe that
they still give a consistent and informative 
picture of the invasion process.

The dynamical exponent $z$ is defined as follows:
Consider an AP initiated at some site following its growth activating
at each time step all the available sites at that moment. The lateral
size of the AP grows as a power law in time
\begin{equation}
t \sim r^z.
\end{equation}
It is easy to show that the fractal dimension of the set of activated sites
at any instant is 
\begin{equation}
D_{ac} = d+\chi_c-z
\end{equation}

Consider the ongoing process of the activations on the surface of the cluster. 
Since the fractal dimension $D_{ac}$ is smaller than $d$, (since $\chi_c<1<z$) 
the set of active sites is so sparse that the time interval
between subsequent activation of a single site will be very long (in fact we
can show that it scales as $r^{z\chi_c/(z-1)}$). This indicates that the
dynamical process of the AP growth is a percolative process, similar
to the process 
the dynamical invasion into a percolation cluster. 
The first conclusion from this argument is that $z=z_{per}$, the
dynamical exponent of invasion percolation. 
This relation has been suggested in \cite{per} and verified
numerically in dimensions $d\le6$.

The AP dynamics in fact resembles a set of invasion percolation
processes occuring simultaneously, so that $D_{ac}$ may be larger than 
the dimension of the set of active sites in a percolation process
$D_{per}-z_{per}$. 
However $D_{ac}$ cannot be larger then the fractal dimension 
of  a full percolation cluster $D_{per}$ without completely destroying
the dynamical picture. Hence consistency demands that
\eqs
D_{ac} = d+\chi_c-z\le D_{per},
\en
and thus,
 \eq{boundchi}
\chi_c\le D_{per}+z_{per}-d\le6-d.
\en
In the last inequality we used the largest possible values for
$D_{per}$ and $z_{per}$ which are the mean field values.
The inequality (\ref{boundchi}) gives an upper bound for the AP
roughness $\chi_c$ in each dimension.
We verified that this bound indeed holds for the numerical values of
$\chi_c$, and is actually rather sharp (see table 1 in subsection 2.5).
Moreover, (\ref{boundchi}) implies that when $d\ge6$, $\chi_c$ must be
0, so we have
 \begin{equation}
d_c\le 6
\end{equation}

We note that for $d>d_c$ our arguments become irrelevant since the AP
is then a fractal objects and its growth in this case will be
occur mostly in the boundary and not in on its  bulk. Therefore we
expect that the process is no longer a percolative one.
It is amusing to note that for this model only the low dimensional
processes in $d\le d_c=6$ are percolative.

\subsection{Numerical simulations in high dimensions}
We performed numerical simulations of the Buldyrev-Sneppen model and
compared them with  the previous theoretical scenario.
The simulations were of either of two kinds:
Interface growth by constant current on a hyper-cubic lattice with
periodic boundary conditions, and invasion of a single cluster 
in a Cayley tree as an effective infinite dimensional
lattice. 

In each of the finite-dimensional simulations the number of activations
was of the order of $10^{10}$, and thus statistical errors were rather
small. However, in high dimensional systems we were severely
constrained by memory considerations. For example in 6 dimensions,
the highest dimensionality in our simulations, the hyper-cube measured
only 16 sites in each directions. Thus, finite size effects were very
strong in the high dimensional cases, leaving a small scaling range
still observable. The numerical investigations on the Cayley tree are
described in detail in section 3.

We measured directly the exponents $\tau_\parallel$, $\gamma$ and
$D$. These exponents are presented in  
table~1. Since only 2 exponents are independent in each case,
measuring three different exponents provides a consistency check.
The values displayed in table~1 do obey the scaling
relations (\ref{sc-par}) and (\ref{sc-opz}), within numerical error.

The most remarkable feature in table~1 is the monotonic decrease of
$\chi_c=D-d$ as a function of $d$. In 6 dimensions the APs become flat
to within numerical accuracy, indicating that $d_c\sim 6$.
This is in accord with our previous estimate for the value of $d_c$
based on dynamic scaling. Since it has been also confirmed in
\cite{per} that $z=z_{per}$ we believe that our
conjecture about the nature of the dynamical process of invasion is
very plausible.
We also note that in 4 dimensions the
measured $\chi_c$ becomes smaller than $\chi_a$, the annealed
roughness. This implies (cf. eq. (\ref{bounds})) that the interface
roughness $\chi$ is dominated by annealed mechanism. In the next
subsection we discuss this point and analyze in detail its consequences.

Another feature evident from table~1 is that in 6 dimensions the
scaling exponents seem well converged toward their infinite
dimensional values.
\begin{table}
\begin{tabular}{|c|c|c|c|c|c|c|}
$d$ & $p_c$    & $\gamma$  & $D$ & $\tau_\parallel$ &
$\tau$ & $\chi$\cite{opz2,shechter}    \\\hline
1  & .4610(2)  &   2.03(3) &  1.64(2)  &   1.42(2)  &      1.25   &  0.67  \\  
2  & .2002(2)  &   1.53(2) &  2.52(2)  &   2.11(2)  &      1.44   &  0.50  \\  
3  & .1148(3)  &   1.29(2) &  3.40(2)  &   2.86(2)  &      1.55   &  0.38  \\
4  & .0785(5)  &   1.21(1) &  4.24(3)  &   3.58(2)  &      1.61   &  0.27  \\
5  & .0583(2)  &   1.13(2) &  5.07(3)  &   4.22(3)  &      1.64   &  0.25  \\
6  & .0445(5)  &   1.08(2) &  6.00(5)  &   5.05(5)  &      1.68   &  0.2   \\
\hline
$\infty$ & .150(1) & 1 & 7.(1) &  6.6(2) & 1.75(5) & 0
\end{tabular}
\caption{Results of numerical simulation of bounded slope models. The
finite $d$ entries refer to simulations on hyper-cubic lattices, and 
the $d=\infty$ entry refers to simulations on a Cayley tree with
coordination number 3. The
exponents $D$, $\tau$ were directly measured, as well as $\gamma$ in
the finite dimensional case. The value of $\tau$ was calculated from
$D$ and $\tau_\parallel$ using (\protect{\ref{sc-par}}), in the finite
dimensional simulations, and measured directly in the Cayley tree
simulation. The value of 
$\chi$ (the overall roughness) was not measured but is given for
reference. 
The numbers in parenthesis are errors in the last digit. We have
checked that the numerical values of exponents presented here satisfy
all the theoretical scaling relations which connect them.}
\end{table}

\subsection{Overall interface roughness}
We now turn our attention to the scaling of the growing interface as
a whole.
Above $d_{c}$ the APs are flat, and the invasion process becomes very
similar to the annealed process: Each AP activates each member of a
set of sites once, or at most a few times.
In this regime the interface may be described 
by the annealed KPZ \cite{gen,kpz}
\eq{an-kpz}\partial_t h({\bf x},t)=\nabla^2 h({\bf x},t)+\lambda 
(\nabla h({\bf x},t))^2+\eta({\bf x},t),\en
where $h$ is the height and the random noise $\eta$ describes the
activation of sites by the flat APs. Since the activation of sites by
APs favors nearby sites, $\eta$ has spatial
correlations derived from the distribution and 
structure of the flat APs. We expect that in large enough dimensions,
the spatial correlations of $\eta$ will become irrelevant, and that
the critical exponents of the interface will identify with those of
KPZ with delta-correlated noise.
We thus define the
{\it interface critical dimension} $d_i$, such that $\chi_a=\chi_q$ for 
$d\geq d_i$. When observing the invasion process of the interface as a
whole, rather than the APs, $d_i$ signifies the dimensionality where
the effects of noise quenching become irrelevant.

It is not difficult to estimate the statistics of the correlated noise
$\eta$, when $d>d_c$, in terms of the AP properties. Since
the APs are not correlated (for $f_0=f_c$, see \cite{opz1}), $\eta$ is
Gaussian, and its value at two points is correlated only if both sites were
activated in the same AP.  Therefore the correlation function may be
estimated by the probability of such an event,
\eq{corr}\begin{array}{r} \left<\eta({\bf x})\eta({\bf y})\right>\sim
P(r_\parallel\geq|{\bf y}-{\bf x}|) |{\bf y}-{\bf x}|^{D-d}\\\sim
|{\bf y}-{\bf x}|^{D(2-\tau)-d}.\end{array} \en

We now analyze the effects of those correlations, using $k$-space
representation of the problem. We make the following definitions:
Let 
\eqs h({\bf k},\omega)=
\int d^d{\bf x}dt\,h({\bf x},t)e^{i{\bf k}\cdot{\bf x}+i\omega t},\en
$\eta({\bf k},\omega)$ defined similarly.
The response function is
\eqs
G({\bf k},\omega)\delta({\bf k}+{\bf k'})\delta(\omega+\omega')=
\left<\delta h({\bf k},\omega)/\delta\eta({\bf k'},\omega')\right>\en
and we assume it has the scaling form
\eqs G({\bf k},\omega)=k^{-z}g(\omega/k^z),\en 
where $z$ is the dynamical exponent of annealed growth.
The noise correlation in $k$-space is
\eqs \left<\eta({\bf k},\omega)\eta({\bf k'},\omega')\right>=k^\alpha
\delta({\bf k}+{\bf k'})\delta(\omega+\omega'),\en
where $\alpha=D(\tau-2)$.

The roughness of an interface described by eq.~(\ref{an-kpz}) is generated
by a combination of the noise correlations and the intrinsic
roughness stemming from the nonlinear process. It is possible to give
a naive calculation for the correlation induced roughness which is a
lower bound to the actual roughness.
To this end we define the ``bare'' height field as
\eqs h_0({\bf k},\omega)=G({\bf k},\omega)\eta({\bf k},\omega) \en
The bare correlation function is
\eq{bare}
\begin{array}{r}\left<h_0({\bf k},\omega)h_0({\bf k'},\omega')\right>
\!=\!G({\bf k},\omega)G({\bf k'},\omega')
\left<\eta({\bf k},\omega)\eta({\bf k'},\omega') 
\right>\\\sim
k^{-2z+\alpha}g(\omega/k^z)^2\delta({\bf k}+{\bf k'})\delta(\omega+\omega'),
\end{array}\en
and the simultaneous bare correlation is given by
\eq{baresim}\left<h_0({\bf k},t)h_0({\bf k'},t)\right>=
 k^{-2z+\alpha}\int d\omega g(\omega/k^z)^2\sim k^{-z+\alpha},\en
so we get the bare roughness
\eqs 2\chi_0=\max(z-\alpha-d,0).\en
Thus, the correlation induced roughness dominates the intrinsic
roughness unless
\eq{dom-sc} 2\chi>z-\alpha-d. \en
One can check that the condition for irrelevance (\ref{dom-sc}) is
always satisfied for uncorrelated noise
($\alpha=0$), and that it also reproduces the known
condition for irrelevance 
of the noise correlation in 1 dimension, $\alpha>-1/2$ \cite{gen}.

The condition for irrelevance (\ref{dom-sc}) was derived for $d\geq d_c$, 
and is satisfied in this range, see table~1 and \cite{kim-kos} for
values of $\chi$ and $z$ for high dimensional KPZ; we can thus conclude that
$d_i\leq6$. The
measured value of $\chi_c$ becomes equal to numerical measurements of $\chi_a$
 at 4 dimensions within numerical accuracy, so $d_i\geq4$.
Since $\chi_c\ll1$ in 5 dimensions, it is probably
safe to use the criterion (\ref{dom-sc}) also for $d=5$. The criterion
is satisfied so that we get the final estimate
\eqs4\leq d_i\leq5.\en
Independent measurements of the overall interface 
roughness\cite{shechter} indicate that this is indeed the case, see
table~I. We conclude that above 4 dimensions the roughness exponent of
the overall quenched growth becomes identical with the annealed
growth exponent.

\section{Cayley tree}

In many statistical physics problems, the infinite dimensional limit
is obtained by defining the model on the Cayley tree. In problems with
an upper critical dimension the Cayley tree solution gives the mean
field exponents. 

The recursive structure of the Cayley tree simplifies
the problems because of lack of self interaction, and allows in many
cases to obtain a closed form solution. 
A well known example is the solution of percolation on the Cayley tree
\cite{bunde}. This solution is instructive since it reveals some of
the characteristics of the percolation clusters.
We are interested in the analogous problem for interface growth.
In this case, however, it will be apparent that 
Cayley tree realizations of this model  might reveals properties which
are sub-dominant for
any finite dimensional realization of the model. We shall show that
strong fluctuations may potentially destroy the usual kind of phase transition
and may lead to anomalous critical properties. 
The Cayley tree realization which we define here 
displays a simple critical behavior with a strong  background of
anomalous behavior, but in other examples the anomalous behavior dominates.

We define the invasion process on the Cayley tree as follows:
The interface is defined by assigning a height to
each site of the Cayley tree. As in the finite dimensional case, there
is a quenched random force $f$ which depends on the site in question
and the height of the interface. A local slope is defined between
every pair of neighboring sites, and the bounded slope condition
prevents the local slope from increasing beyond a critical value by
advancing the sites which violate this condition. 
As usual there is a critical value of force $f_c$, above which there
is as finite probability to create an infinite cluster.

The avalanche propagates upwards and outwards from the original site. 
Since we expect critical avalanches to be flat, propagation of
avalanches will be mainly in the outward direction.
However, two additional processes complicate the dynamics. First, the
bounded slope rule implies that repeated activation of a single site
is followed by activation of all its near neighbors, and the number
of these neighbors grows exponentially with the number of repeated
activations of the original site. Second, since the activations are
isotropic, sites may be activated, create sub-avalanches, and then be
reactivated by a sub-avalanche created by of their children. Such
backward activations creates self-interaction in the dynamics, and is
one of the main reasons for the difficulty in analyzing the model in a
quantitative way. This self-interaction is
the reason that the model on a 1-dimensional chain,
being a special case of the Cayley tree, is not trivially solvable.

The general origin of the anomalous critical behavior in the size
distribution of the clusters on the Cayley was discussed briefly in
the previous letter \cite{letter}. We will give an example below,
where this mechanism can be studied analytically, but first we would
like to present the qualitative arguments, since they have an
essential role in the behavior of the phase transition.

Consider an interface that is initially flat, and a growth process 
initiated from a
single site with a fixed value of $f_0$.
There is a finite probability
that the same site will be activated once more, activating the $q$
nearest neighbors obeying the bounded rule. $h$ subsequent activations
of the same site will result in activation of the
$q(q-1)^h$ neighboring 
sites, again as a result of the bounded slope condition.
We observe that the number of activations in this sub-process
grows exponentially with $h$. However, since the probability for each
of the activation is $f_0$, the probability for such an event is
$f_0^h$ which is exponentially small. 
Thus, even
though the probability to create such a height difference decreases
exponentially with $h$, the effect might become  important because of the
exponential avalanches that follow from such events.
This effect can dominate the probability distribution of~$s$.

The actual number of sites which are activated by the original site of
height $h$ is larger than the previous estimate, since other
sites can be activated by the exponential sub-avalanche. Another
complication arises since the actual interface is not necessarily
flat. However, since it is statistically flat, a sufficiently long
sequence of activations should have an exponential effect. Therefore
we estimate  the mean number of activations ${\bar s}(h)$ resulting from the
event that a single site is of height $h$, by 
\eqs {\bar s}(h)\sim\mu^h,\en
with $\mu\ge q-1$. 
On the other hand, the probability that a site in an $f_0$ avalanche
reaches a height $h$ should decrease like ${\tilde f}^h$ with $\tilde f$
larger than, but of the order of $f_0$. We can combine these two
estimates to find the probability of an avalanche of size $s$,
\eqs {\sf P}[{\bar s}(h)\leq s\leq{\bar s}(h+1)]\sim {\tilde f}^{h}; \en
it follows that
\eq{sc-bte} P(s)\sim s^{(\log{\tilde f}/\log\mu)-1}\equiv s^{\tau(f_0)}. \en
Thus on the Cayley tree, the tail of the distribution of
the APs sizes is always a power law, with a non-universal $f_0$-dependent
exponent.

Near the phase transition there are two possible scenarios. Either the
transition large clusters are created by the present mechanism, or
they are the result of critical fluctuations. In the first case
there is no typical scale for the mass of the clusters, as follows
from (\ref{sc-bte}). However
there is a typical {\it length} scale which remains finite in the
transition. 

The second case may be realized if for sufficiently small
$|f_0-f_c|$, the probability to create a  very large critical cluster
is larger than the probability displayed in (\ref{sc-bte}). Thus, there will
be a crossover scale, beyond which normal critical behavior
dominates. In particular, there will be typical scales for the
mass and (chemical) length of the clusters, which diverge at $f_c$, and the
clusters will have a finite fractal dimension. Since the mechanism which
creates anomalous clusters is not universal, the question which
critical behavior is realized depends on the specific realization.

In the following subsections we are going to discuss, in addition to
the model which we have already presented, two additional realizations
of the bounded slope model on the  Cayley tree. One realization was
defined and analyzed in \cite{shechter}, and in it the interface grows
into the tree, instead of in an additional direction. The other
realization, to be defined in detail below, is a directed version of
the model defined above, where backward activations are
disabled. These two alternative realizations are integrable, and
display anomalous critical behavior. The original model, however,
displays numerical evidence for crossover to standard critical behavior.

\subsection{Numerical analysis on the Cayley tree}
We performed Monte-Carlo simulations of the bounded slope model on the
Cayley tree with coordination number~3. We used the constant driving
force algorithm version, {\it i.e.}, for each site which was a candidate
for activation, a random number between 0 and 1 was compared with a
constant force $f_0$, to decide whether the site should be activated. This
process was repeated until either the cluster was blocked, or exceeded
an upper bound in size.

We collected the statistics of clusters, in terms the cluster
size $s$, and in terms of maximal chemical distance from the source
site $\ell$. In Fig. 1 we give examples of
the size distribution for several values of $f_0$. We observe three
qualitative types of behavior, which can be explained in terms of the
interplay between critical and anomalous fluctuations described above.
In particular, the size distribution very near to the critical value
of forcing displays a crossover behavior, which we interpret as
emergence of normal critical behavior.

\begin{figure}
\epsfxsize=8truecm
\epsfbox{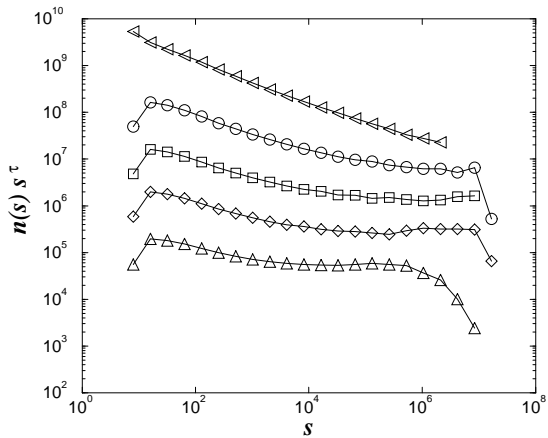}
\label{Figure 1}
\caption{The distribution of cluster sizes $n(s)$ multiplied by the
critical distribution $s^{\tau_c}$ (arbitrary units), given as a function
of $s$. We present distributions for forcing levels
$f_0=0.141,\,0.145,\,0.148,\,0.150\sim 
p_c$ and $0.155$, displayed with higher values of $f$ below lower
ones. For low values of $f$, below $p_c$, one can observe 
pure anomalous power law distributions, while as $f$ increases a new
feature becomes more and more apparent for large $s$, as one
approaches $p_c$. The new power law seems to be independent of $f$
near $p_c$. Above $p_c$ an exponential cutoff appears for large $s$.}
\end{figure}

The distribution functions for subcritical driving force display a
clear power law tail without a cutoff, but with an $f_0$
dependent exponent $\tau(f_0)$ (see eq. (\ref{sc-bte})). The clusters in
the tails of the subcritical 
distributions were created by the mechanism described in the last
section. An unusual property which follows from the asymptotic form of
the subcritical distributions, is the divergence of the average
cluster size $\left<s\right>$ when $\tau<2$. This property has no
significance beyond the fact that the size distributions have no cutoff
in the subcritical range.

The supercritical size distributions, on the other hand, display well
defined cutoffs. The presence of these cutoffs stems from exactly the
same reasons as the supercritical cutoff in usual critical phenomena;
once a cluster reaches a certain size, it becomes very unlikely that
it will not become an infinite cluster, and this is true regardless
of the of mechanism which creates the power law tails in the
subcritical range.

Very close to $p_c$ ($f_0=0.150$) a third type of distribution is
observed. There is sub-asymptotic power law tail with an exponent
$\tau(f_0)$ characteristic of the subcritical distributions, but at a
certain value of $s$ there is a crossover to a different power law
distribution with an exponent $\tau_c=1.75\pm0.05$. 
We attribute the crossover to
the emergence of normal critical behavior, which is masked by the
anomalous fluctuations for other values of $f_0$. The fact that very
large clusters near $f_c$ are generated mainly by critical
fluctuations signifies that the transition itself is of normal type.

Analysis of the length distributions of clusters $p(\ell)$, displayed in
fig. 2, supports the picture obtained from the analysis of mass
distribution. The
subcritical distributions decay faster than a power law, but
for $f_0\sim f_c$ the distribution decays asymptotically as a power law
with a characteristic exponent $\tau_\ell=3.8\pm0.1$, consistent with
normal critical behavior at $f_c$.
\begin{figure}
\epsfxsize=8truecm
\epsfbox{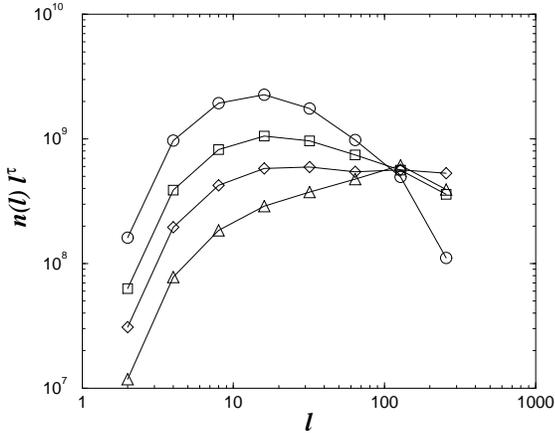}
\label{Fig 2}
\caption{The distribution of clusters according to their chemical
length $\ell$, multiplied by the critical distribution
$\ell^{\tau_\ell}$, for values of $f$ corresponding to those displayed
in fig. 1, except the smallest value. One can observe clearly that the
distribution falls faster than a power for $f<p_c$, and is power like
at $p_c$. }
\end{figure}

We conclude that depinning transition of the bounded slope model on
the Cayley tree is a critical phenomenon with a very strong background
of anomalous fluctuations. Using the critical exponents $\tau_c$ and
$\tau_\ell$ one may also calculate the fractal
dimension of critical the clusters
$D=2(\tau_\ell-1)/(\tau_c-1)\sim7$. The large 
value of $D$ shows that this phase transition is not in the
percolation transition universality class. The critical exponents are
also listed for reference in table~1.

\subsection{A solvable model}
It was remarked above that the main obstacle for obtaining an exact
solution for the invasion model on the Cayley tree is self-interaction
through reactivation of a site by a ``child'' site. A natural way to
overcome this difficulty is to define a model in which there is a
preferred direction, `back', in which the bounded slope rule is not
enforced. The mechanism which creates anomalous large fluctuations is
present in this simplified directed model, and a more precise analysis of this
feature is possible. However, in contrast with the ``isotropic''
Cayley tree model which was discussed in the previous subsection, we
show that in
the directed model anomalous fluctuations dominate up to the 
transition, and normal critical behavior is not observed. In addition
to the directed model, we also discuss briefly the model presented in
\cite{shechter}, and reach similar conclusions.

The directed model is therefore defined by precisely the same rules as
the original model defined above, except that the slope may be
arbitrarily large when the lower point is closer to the root.
Thus defined, the
process can be carried out also in the following manner:
Pick a root site on the tree and let it be activated repeatedly
until further activation is blocked by a pinning force larger than
some fixed driving force $p_0$. The
probability that a height
$h$ will be reached by this process is $p_0^h(1-p_0)$. Since the
bounded slope rule works only forward, the interface height determined
by this process may not be changed anymore, and is statistically
independent of the interface height at any other site.
Next, the same process is carried out for each of the child sites of
the root, starting from height $h-1$ if $h\ge2$; otherwise the
invasion process stops. The process continues analogously for each of
the children sites either until all the sites are blocked, or indefinitely,
creating an infinite cluster (this can happen only when $p_0>p_c$). 

The present model is
equivalent to a branching process~\cite{harris}, and we may use its recursive
structure for analysis, similar to the analysis of Cayley tree
percolation. However,  the
model defines a universality class different from percolation,
since it is a branching process with an infinite number of states:
Each possible value of the interface height corresponds to a
different branching process state.

The mass distribution of clusters of the directed model may be
calculated as follows. We 
denote the probability that the total mass of a cluster is $s$ given
that the interface height at the root site is $h$ by $p(s|h)$. This function
also describes the probability that, given a child site with interface
height $h$, it will generate a sub-cluster of mass $s$. Since the total
mass of the cluster is equal to the sum of the masses of all the
sub-clusters plus the number of activations at the root, we have
\eq{srecur}\begin{array}{r} 
p(s|h)=\displaystyle\sum_{s_1+\dots+s_q+h=s}\sum_{h_1\ldots h_q}
p(s_1|h_1)p(h_1|h)\quad\\\times p(s_2|h_2)p(h_2|h)\ldots
p(s_q|h_q)p(h_q|h).\end{array}\en
The RHS of eq. (\ref{srecur}) is a sum of products of expressions of the form
\eq{fp}\begin{array}{l}  f(s|h-1)\equiv\displaystyle\sum_{h'}
p(s|h')p(h'|h)\\\qquad =\displaystyle\sum_{h'=0}^\infty
(1-p_0)p_0^{h'}p(s|h'+h-1). \end{array}\en
It is possible to invert (\ref{fp}) to express $p(s|h)$ in terms of
$f$,  and rewrite eq. (\ref{srecur}) 
\eq{srq}\begin{array}{l} 
(1-p_0)^{-1}(-p_0f(s|h+1)+f(s|h))\quad\\\displaystyle=
\sum_{s_1+\dots+s_q+h=s}
f(s_1|h-1)f(s_2|h-1)\ldots f(s_q|h-1). \end{array}\en
The Laplace transform of the last equation is
\eq{srl}(1-p_0)^{-1}(-p_0g(x|h+1)+g(x|h))=x^hg(x|h-1)^q, \en
where $g(x|h)\equiv\sum_s x^s f(s|h)$ is the Laplace transform of $f(s|h)$.
The solution $g$ of this non-linear difference equation contains all
the information about the mass distribution. In particular many
properties of the distribution $p(s|h)$ follow from the
behavior of $g(x|h)$ near $x=1$. For example, the probability to create
an infinite cluster is
\eqs p_\infty(h)=1-{g(1|h)-p_0g(1|h+1)\over1-p_0}.\en 
Similarly, the expected cluster size starting at height $h$ is
\eq{expval}\left<s|h\right>={\partial_x g(1|h)-p_0\partial_x g(1|h+1)
\over 1-p_0}.\en
$\left<s|h\right>$ obeys a linear difference equation obtained by linearizing
(\ref{srl}) near $g(1|h)$. For small enough $p_0$, $g(1|h)$ must be 1,
and we can linearize eq.~(\ref{srl}) near $x=1$ to get,
\eqs {-p_0\partial_xg(1|h+1)+\partial_xg(1|h)\over(1-p_0)}=q\partial_x
g(1|h-1)+h,\en
which can be solved in closed form, yielding expressions for
$\left<s|h\right>$. The solution 
ceases to be realizable when $p_0=p_c$, where $p_c$ is defined by
\eq{pc} 2qp_c(1-p_c)=1.\en
This implies that for $p_0>p_c$, $g(1|h)\ne1$, so that $p_c$ indeed is
the critical value of $p_0$.

Further properties may obtained by solving eq. (\ref{srl})
numerically. One finds that for $p_0$ near $p_c$ and $1-x$ small
\eq{gtaylor} g(x|h)\sim 1+(1-x)g_1(h)+g_s(h)(1-x)^\tau,\en
where $2<\tau<3$. This type of asymptotic behavior implies, using the 
Tauberian theorems \cite{shechter}\cite{feller},
that $f(s|h)$ and therefore also $p(s|h)$ behave
asymptotically as $s^{-\tau}$ for large $s$, and in particular that
$\left<s|h\right>$ is finite but  $\left<s^2|h\right>$ diverges for
$p_0\le p_c$ (near $p_c$). 
For example, for $q=2$, $p_0=p_c$ we find that $\tau=2.2\pm0.05$

We observe that the subcritical mass distribution of the directed
model is quite similar to that of the original model, with a 
$p_0$-dependent power law tail, but in contrast with the original model, 
this behavior persists for values of $p_0$ up to and including $p_c$.
This indicates that the directed model belongs to the
second case outlined in the beginning of this  section, where the
phase transition
is dominated by the anomalous fluctuations caused by exponentially large
compact avalanches. The same conclusion also arises from the analysis of
length distribution of the clusters, presented next.

The analysis of cluster length is more involved than that of cluster
mass. Using the analogy with an infinite state branching process, we
interpret the sites at a given chemical distance $\ell$ from the root,
as the $\ell$th generation of the process. The population of the
process is defined by the values of the interface height, where each
height corresponds to a different species of the branching
process. Thus, we may define the $\ell$th generation population vector
${\bf Z}_\ell=(Z_\ell^{(1)},Z_\ell^{(2)},....)$, where $Z_\ell^{(1)}$
gives the number of sites at chemical distance $\ell$ from the root
with interface height equal to 1, etc.  The branching process is
defined by giving the probability $p_{n}({\bf Z})$ that an individual
of type $n$ will generate a set ${\bf Z}=(Z^{(1)},Z^{(2)},....)$ of
individuals in the next generation.  The values of these probabilities
follow from the definition of the directed model.  Since we were
mainly interested in the asymptotics of the length distribution of
clusters, we analyzed the dependence of the moments
$\left<Z_{\ell}^{(n)}\right>$ and
$\left<Z_{\ell}^{(n)}Z_{\ell}^{(m)}\right>$ on $\ell$.  Such moments
average both over clusters shorter than $\ell$, contributing 0, and
over clusters longer or equal to $\ell$. For critical clusters we
expect that the distribution of ${\bf Z}_\ell$ does not depend on the
cluster size, given that cluster is longer or equal to $\ell$. Thus,
we may write
\eq{fm} \left<Z_{\ell}^{(n)}\right> = p(\ell)
\left<Z_{\ell}^{(n)}|{\bf Z}_\ell\ne0\right>
\sim c_n p(\ell) s(\ell),\en
where $p(\ell)$ gives the probability that a cluster is longer than
$\ell$, and $s(\ell)$ 
is the average number of sites at a distance $\ell$ from the
root. $s(\ell)$ would be 
of the form $\ell^{D_\ell-1}$ if the critical clusters have a finite
fractal dimension $D_\ell$. Similarly 
\eq{sm} \left<Z_{\ell}^{(n)}Z_{\ell}^{(m)}\right> \sim c_{n,m} p(\ell)
s(\ell)^2,\en   
so that we may deduce the behaviors of $p(\ell)$ and $s(\ell)$ from those of
the first and second moments of $\bf Z$.

Explicit calculations, not reproduced here, show that first
moments behave as a power law for large $\ell$, while second moments grow
exponentially. This implies, using (\ref{fm}) and (\ref{sm}), that
$s(\ell)$ grows exponentially and $p(\ell)$ decays exponentially with
$\ell$. In other words, the clusters have a characteristic length, and
have an infinite fractal dimension. These 
properties are expected for a depinning transition dominated by anomalous
fluctuations, as was already indicated by the analysis of cluster mass
distribution. We conclude that in the directed model critical
fluctuation are too 
weak to be observed even near~$p_c$.

An alternative version for defining the bounded slope model on the
Cayley tree was given by \cite{shechter}. In this version, the Cayley
tree is divided into sets of sites with a fixed height, and the
invasion process occurs directly from one site to its neighbor, rather
than by raising the interface in a single site as in the version
presented in the present paper. The paper \cite{shechter} presents an
analysis of the cluster mass distribution. As in the models presented
here, the mass distribution displays non-universal power law
asymptotics, stemming from anomalous fluctuations. Since there is no
indication of a crossover to a different type of asymptotics near
$p_c$, we conclude that in this model the transition is anomalous.

\subsection{Consequences for large but finite $d$}

In many cases one uses information from Cayley tree realizations of
statistical models to deduce properties of high-dimensional realizations.
In particular, one may hope to find ``mean field'' behavior which should
be valid above a finite critical dimension $d_c$, if such exists. 

In the case of the bounded slope models, however, we have a dilemma:
The behavior on the Cayley tree is complicated by the creation of
exponentially large avalanches so we observe two types of critical
behavior. In some cases we had a fluctuation dominated phase transition
and in others the phase transition is of standard critical type.

We believe that the properties observed on the symmetric Cayley tree
should serve as a better guide to the finite dimensional case, for 
two reasons.  First, the symmetric model we used is closer in its
definition to high dimensional bounded slope models than both the
directed model we solved and the tree model suggested in \cite{shechter},
since it has a definite height and symmetric activations.

The second reason follows by examining the process which creates the
anomalous avalanches in large but finite $d$.
The analogous arguments
which led to the conclusion that the cluster mass distribution has power
law asymptotics (cf. eq. \ref{sc-bte}) repeated in the case of finite
$d$ yield a
distribution which behaves asymptotically as a stretched exponential
$\exp(-s^{1/d})$. Such behavior will always become sub-dominant for $p_0$
very close to $p_c$. Thus in a finite dimension we
always expect a crossover from an anomalous
fluctuation regime, to a critical regime, such as displayed by the
symmetric Cayley tree model.
We thus conclude that in large dimensions
the associated processes become objects with a fractal dimension
approaching $D_\infty=7\pm1$. Since in six dimensions $D$, as well as
other critical exponents are close to their infinite dimensional
values, we are led to conjecture that there exists a finite critical
dimension beyond which infinite-dimensional critical behavior is
achieved.

\section{Conclusions} 
The analysis presented in this paper allows us to draw conclusions on
the processes belonging to the quenched KPZ universality class and
processes of percolation of directed surfaces.
We believe that studying the behavior in different dimensionalities
has provided a lot of interesting insights. The main conclusions and
ideas presented in this paper in our view are as follows:

There exist two different roughness exponents characterizing the
invading front: The exponent $\chi_c$ of the AP roughness, and
the overall interface roughness $\chi_q$. The overall roughness is
created by random deposition
of objects with a roughness $\chi_c$. We are in fact able to predict
the exponent $\chi_q$ for $d > 4$ using this understanding.
A basic conclusion of this is that different driving methods
of this model will display different statistical properties.

Another basic idea is that the dynamic process of avalanche formation
below 6 dimensions is a percolative 
process. There are two main conclusions from this claim. Firstly, the
roughness exponent $\chi_c$ goes down with increasing dimension, and the
dimension where $\chi_c = 0$ is $d_c=6$. 
Secondly, the dynamic exponents of the process should be the same as
in percolation \cite {per}.

From our analysis on the Cayley tree we find that in high
dimensions strong fluctuations appear as a result of exponentially rare
events, as well as a result of critical behavior. These fluctuations are
observed by the presence of wide tails in the AP size distribution,
and by power law tails in the cluster distribution on the Cayley
tree. As a result, one can observe two types of 
scenarios for the depinning transition on the Cayley tree. In addition
to a critical transition, it is also possible to have a fluctuation
dominated transition, characterized by a correlation length which does not
diverge. Indeed, this behavior is observed both
in a directed model we solve and in a model analyzed in
\cite{shechter}. On 
the hand, on a simple symmetric bounded slope model the usual critical
transition is realized. We conclude that the latter model is the
appropriate high dimensional limit.

Examination of the values of scaling exponents obtained numerically
(see table~1) indicates that there is a finite absolute critical dimension,
in which all the exponents reach their infinite dimensional limit.

In summary, we find that there are at least three qualitative transitions
in the basic structure of an invading interface when dimensionality is
increased. This reflects the wealth of processes occuring during such
an invasion process.

\end{document}